\documentclass[preprint,number,sort&compress,twocolumn,3p]{elsarticle}
\pdfoutput=1
\usepackage{rotating}
\usepackage{graphicx}
\usepackage{array}
\usepackage{amsmath}
\usepackage{color}
\usepackage{amssymb}
\usepackage{tabularx}
\usepackage{slashed}
\usepackage[pdftex,hyperfootnotes=false,pdfpagelabels]{hyperref}

\usepackage{tikz}
\usetikzlibrary{arrows,decorations.pathmorphing,backgrounds,positioning,fit,petri,automata,shadows,calendar,mindmap,
decorations.markings,calc}

\def\bea{\begin{eqnarray}}
\def\eea{\end{eqnarray}}

\newcommand{\nn}{\nonumber}

\newcommand{\be}{\begin{equation}}
\newcommand{\ee}{\end{equation}}

\newcommand{\bfr}{\begin{mdframed}[backgroundcolor=gray!20] }
\newcommand{\efr}{\end{mdframed}}

\begin{document}

\title{Deciphering the CP nature of the 750\,GeV resonance\tnoteref{t1}}

\tnotetext[t1]{This article is registered under preprint number: DESY 16-054, arXiv:1604.02029.}

\author{Mikael Chala$^{a}$\fnref{fn1}}
\author{Christophe Grojean$^{a,b}$}
\author{Marc Riembau$^{a,c}$}
\author{Thibaud Vantalon$^{a,c}$}

\fntext[fn1]{mikael.chala@desy.de.}

\address{${}^{a}$ DESY, Notkestra\ss e 85, D-22607 Hamburg, Germany}
\address{${}^{b}$ on leave from ICREA, E-08010 Barcelona, Spain}
\address{${}^{c}$ IFAE, Barcelona Institute of Science and Technology (BIST) Campus UAB, E-08193 Bellaterra, Spain}

\begin{abstract}
The recently observed excess in diphoton events at around 750\,GeV can be satisfactorily described in terms of a new spin-0 real singlet with effective interactions to the gauge bosons. In this letter we first review the current constraints on this setup. We further explore the production in association with a gauge boson. We show the potential of this channel to unravel current flat directions in the allowed parameter space.
We then study the potential of two different asymmetries for disentangling the CP nature of such a singlet in both gluon fusion and vector-boson fusion. For this matter, we perform an estimation of the efficiency for selecting signal and background events in eight different decay modes, namely $4\ell, 2j\, 2\ell, 2j\ell\,\slashed{E}_T, 2\gamma\,2j, 4\ell\,2j, 2\ell\,\gamma\,2j, 4j\,2\ell$ and $4j\,\ell\,\slashed{E}_T$. We emphasize that the very different couplings of this new singlet to the Standard Model particles as well as the larger mass provide a distinctive phenomenology with respect to Higgs searches. We finally show that a large region of the parameter space could be tested within the current LHC run, the dominant channel being $2\gamma\, 2j$.
\end{abstract}

\maketitle

\section{Introduction}
The first bunch of data in proton-proton collisions at $\sqrt{s} = 13$\,TeV were successfully delivered by the LHC during last year. Surprisingly, the first analyses on these data with as few as $\sim 3$\,fb$^{-1}$ have revealed unexpected results. Indeed, the ATLAS~\cite{ATLAS:2015dp} and CMS~\cite{CMS:2015dxe} experiments have pointed out an excess in diphoton events with an invariant mass of around 750\,GeV; the local significance ranging from 2 to around 3$\sigma$. The reported excess survived further scrutiny~\cite{MOR:2015} and appears as the best hint in decades for physics beyond the Standard Model (SM) at colliders. 
This fact explains the excitement of the particle physics community, which has translated into a plethora of papers in a few months~\cite{all:2015}.

This diphoton excess can be easily interpreted  in terms of a spin-0 real singlet (although explanations in terms of spin-1 and spin-2 particles have also received some well deserved attention). Both production and decay are hence to be mediated by heavier states\footnote{The model proposed in~\cite{Kats:2016kuz} is an exception where the diphoton excess originates from a solitary new degree of freedom without the need for any additional electrically charged particles, nor new strong dynamics. Alternative non-resonant models with long decay chains have also been proposed to {\it explain} the 750\,GeV diphoton excess. In this letter, we limit ourselves to the {\it simplest} interpretation with a single resonance whose couplings to gluons and photons are mediated by additional heavier states charged under QCD and QED.} whose effects can be encoded in a small set of effective operators. Throughout this letter we adopt this approach and we explore the potential of the next run of data to unravel the CP nature of this candidate, namely whether it is a scalar or a pseudo-scalar (of course, the 750\,GeV singlet could also be an admixture of both scalar and pseudo-scalar, i.e. it could have some interactions that are not invariant under CP transformations, but we omit this possibility for now.) We start considering a generic parameterization in section~\ref{sec:par} and discussing the current constraints. Production via gluon fusion (GF) and vector-boson fusion (VBF) turn out to be sizable in a large region of the parameter space. However, they are shown to give flat directions that can be only disentangled if new production mechanisms are considered. In this respect, we explore the potential of producing the singlet resonance in association with a Standard Model (SM) gauge boson in section~\ref{sec:associated}.
The rest of the article is structured as follows. In sections~\ref{sec:gf} and \ref{sec:vbf}, we introduce two asymmetries in the kinematical distributions of GF and VBF events. They are intended to differentiate the two CP hypothesis.
The advantage of this approach relies on the fact that most systematic uncertainties cancel out. Statistical uncertainties are on the other hand properly taken into account. We perform  simulations to estimate the efficiency for selecting signal and background events in both categories in eight different decay modes: $4\ell, 2j\, 2\ell, 2j\ell\,\slashed{E}_T, 2\gamma\,2j, 4\ell\,2j, 2\ell\,\gamma\,2j, 4j\,2\ell$ and $4j\,\ell\,\slashed{E}_T$. We show that after all cuts, sizable efficiencies are obtained for most signals while still keeping backgrounds under control. Despite that we do not attempt to optimize these cuts, all together the eight channels can probe a wide region of the available parameter space within the current run of the LHC, as explained in section~\ref{sec:results}. We conclude in section~\ref{sec:conclusions}.

\section{Parameterization and current constraints}
\label{sec:par}
\begin{figure*}[t]
 \hspace{-0.25cm}\includegraphics[width=\columnwidth]{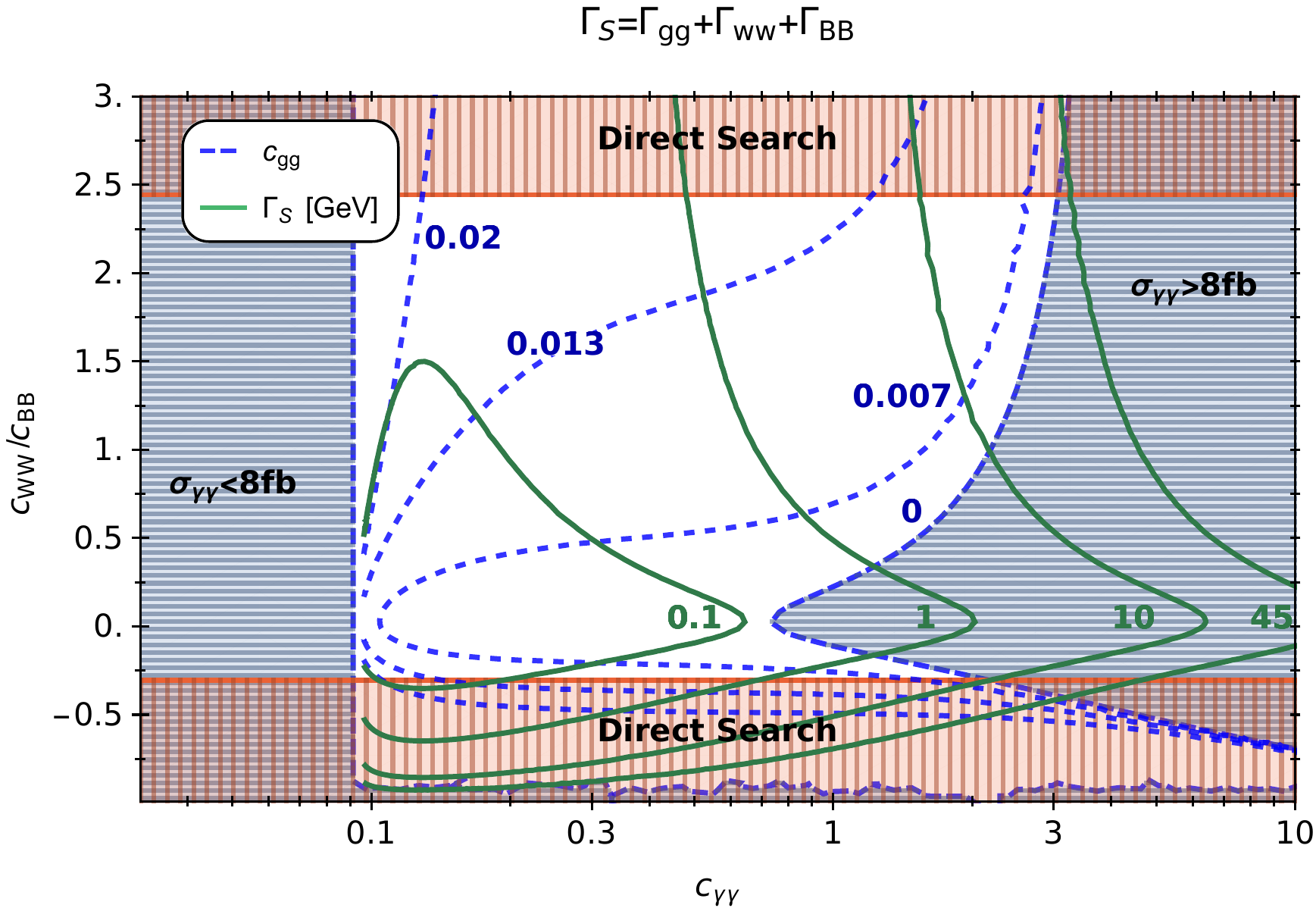}\hspace{0.5cm}
  \includegraphics[width=\columnwidth]{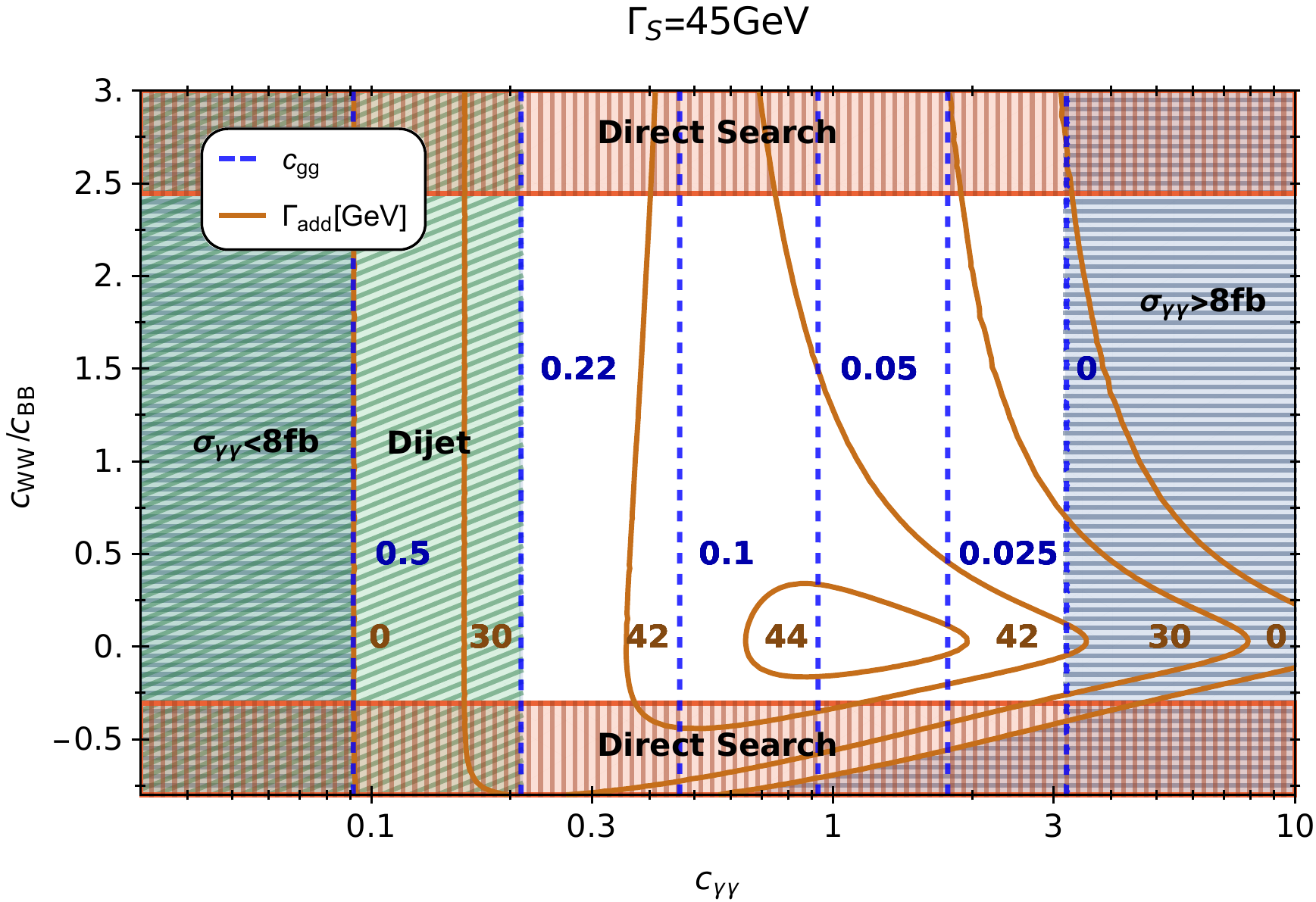}
\caption{Parameter space region in the plane $c_{\gamma\gamma} \,-\, c_{WW}/c_{BB}$ compatible with the diphoton excess and current constraints. Regions filled with horizontal lines can not account for the observed signal. Regions filled with vertical lines are in turn bounded by direct searches at 8\,TeV.  The values of $c_{gg}$ are labeled on dashed contour lines.
In the left (right) panel $\Gamma_S = \Gamma_{gg} + \Gamma_{WW} + \Gamma_{BB}$ ($\Gamma_S = 45$\,GeV) is assumed. The total (additional) width is shown in solid green (brown) lines in the left (right) panel.}
\label{fig:parameterspace}
\end{figure*}
This letter aims mainly to provide a guideline for future efforts on the analysis of the parity properties of a resonance $S$ with mass $M\sim 750$\,GeV. We assume $S$ to be a spin-0 SM gauge singlet. Besides, the production cross section into diphotons mediated by $S$ is assumed to be 8\,fb. The question of the spin and parity properties of $S$ is made legitimate by the unexpected character of the excess and thus by the absence of any particular theoretical prejudice towards one hypothesis.
Actually we do not focus on any particular model nor we attempt to address the effective-field theory of $S$ in full generality. 
In fact, the relevant Lagrangian for our phenomenological study can be parameterized as~\cite{Franceschini:2015kwy}
\begin{align}\label{eq:par}\nn
 \mathcal{L} &=  \frac{1}{2M}S\bigg(g_3^2  c_{gg}  G^2 + g_2^2c_{WW}  W^2 + g_1^2 c_{BB}  B^2 \\
& + g_3^2\tilde{c}_{gg} G\tilde{G} + g_2^2\tilde{c}_{WW} W\tilde{W} + g_1^2\tilde{c}_{BB} B\tilde{B} \bigg).
\end{align}
Here, $g_3, g_2$ and $g_1$ stand for the SM $SU(3), SU(2)$ and $U(1)$ gauge couplings, respectively. $G$, $W$ and $B$ are the corresponding field-strength tensors. For a generic $F$, $\tilde{F}$ is defined as $\tilde{F}_{\mu\nu} = \frac{1}{2}\epsilon_{\mu\nu\alpha\beta}F^{\alpha\beta}$. The tilded (non-tilded) coefficients are zero if $S$ is a scalar (pseudo-scalar). We disregard further couplings to the SM fermions and  to the Higgs doublet since they do not introduce any qualitative change in our analysis. Actually the latter has anyway to be small to pass the constraints from  Higgs measurements~\cite{comb:2015} and $ZZ$ resonant searches~\cite{Aad:2015kna}.
The decay width of $S$ into the different decay modes provided by the interactions above can be easily computed for $M \gg m_{W,Z}$, with $m_{W(Z)}$ the mass of the $W^\pm (Z)$ boson. In this limit, the decay widths to the different pairs of gauge bosons are given by
\begin{eqnarray}\nn
\Gamma_{gg} &=& 8 \pi \alpha_3^2 M \left( c_{gg}^2 + \tilde{c}_{gg}^2 \right),\,\\\nn
\Gamma_{\gamma\gamma} &=& \pi \alpha_{\text{em}}^2 M \left( c_{\gamma\gamma}^2 + \tilde{c}_{\gamma\gamma}^2 \right), \\\nn
\Gamma_{Z\gamma} &=& 2 \pi \alpha_{\text{em}}^2 M \bigg[ \left(c_{BB} t_W - \frac{c_{WW}}{t_W}\right)^2 \\\nn
&&+ \bigg(\tilde{c}_{BB} t_W -\frac{\tilde{c}_{WW}}{t_W}\bigg)^2 \bigg]\,,\\\nn
\Gamma_{ZZ} &=& \pi \alpha_{\text{em}}^2 M \bigg[ \left(c_{BB} t_W^2 + \frac{c_{WW}}{t_W^2}\right)^2 \\\nn
&&+ \left(\tilde{c}_{BB} t_W^2 + \frac{\tilde{c}_{WW}}{t_W^2}\right)^2 \bigg]\,, \\
\Gamma_{WW} &=& \frac{2 \pi\alpha_{\text{em}}^2 }{s_W^4} M \left( c_{WW}^2 + \tilde{c}_{WW}^2 \right)\,,
\end{eqnarray}
with $c_{\gamma\gamma}=c_{BB}+c_{WW}$, $t_W$ and $s_W$ the tangent and sine of the Weinberg angle, $\alpha_\text{em}$ the fine-structure constant and $\alpha_3 = g_3^2/(4\pi)$. The photon field-strength coefficient is thus given by $4\pi\alpha_\text{em} c_{\gamma\gamma}/2M$.
The cross section for the single production of $S$ and the subsequent decay into two photons at a center of mass energy $\sqrt{s}$ reads
\begin{equation}
\sigma^{\gamma\gamma}(s) \,=\, \frac{1}{s}\frac{1}{M \Gamma_S} \left( \mathcal{C}_{gg} \Gamma_{gg} + \mathcal{C}_{\gamma\gamma} \Gamma_{\gamma\gamma} \right) \Gamma_{\gamma\gamma},
\end{equation}
where $\Gamma_S$ stands for the total width. $\mathcal{C}_{gg}$ and $\mathcal{C}_{\gamma\gamma}$ represent instead dimensionless parton luminosities for gluon and photon fusion, respectively. Their values at 8 (13)\,TeV have been found to be approximately 174 (2137) and 11 (54), respectively~\cite{Franceschini:2015kwy}. 
The single production of $S$ via GF at $13$\,TeV is thus enhanced with respect to 8\,TeV by a factor of $\sim 5$, which can be in agreement with the absence of departures from the SM predictions in the first LHC run. This in fact translates into a bound on  $\sigma^{\gamma\gamma}(8\,\text{TeV}) \lesssim 2$\,fb~\cite{CMS:2014onr,Aad:2015mna}. This observation is no longer true for single production via photon fusion. It is only increased by a factor of $\sim 2.9$ and  therefore in tension with current constraints (see for example~\cite{Harland-Lang:2016qjy,Panico:2016ary}).
The $c_{\gamma\gamma}$ coupling is bounded from above (below) to avoid too large (small) a diphoton cross section. In the same vein, experimental searches for resonant $Z\gamma$~\cite{Aad:2014fha}, $ZZ$~\cite{Khachatryan:2015cwa} and $W^+W^-$~\cite{Aad:2015agg} production at 8\,TeV set stringent limits on this setup.
This information is summarized in Fig.~\ref{fig:parameterspace}. The allowed parameter space in the $c_{\gamma\gamma} \,-\, c_{WW}/c_{BB}$ plane that can explain the excess while evading the current bounds is presented in this plot. For every point in this plane, $c_{gg}$ has been fixed so that $\sigma^{\gamma\gamma}(13\,\text{TeV}) = 8$\,fb. The corresponding values are shown in dashed blue lines. The region filled with vertical lines is excluded mainly by $W^+ W^-$ and $Z$ searches at 8 TeV. Notice also that the bounds coming from direct searches would be  weaker if $\sigma^{\gamma\gamma}(13\,\text{TeV})$ was smaller. The solid green (brown) contour lines stand for the total (additional) width. In the left panel of the figure we assume that $\Gamma_S$ coincides with these contours. In the right panel we fix it instead to the best-fit value reported by the ATLAS Collaboration~\cite{ATLAS:2015dp}, $\Gamma_S = 45$\,GeV, by considering an additional partial width of $S$ into soft (or partially invisible) particles that escape detection.

\section{Associated production}
\label{sec:associated}
\begin{figure}[t]
\hspace{0.3cm}\centerline{\includegraphics[width=1.1\columnwidth]{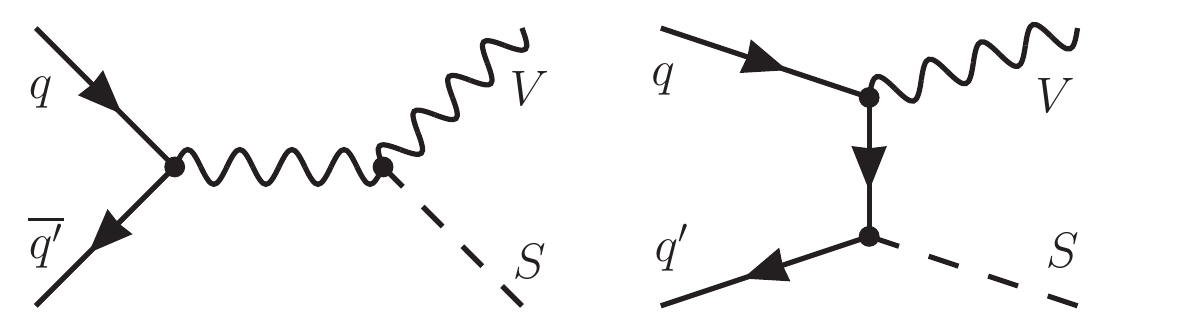}}
\caption{Feynman diagrams for $S$ production in association with a gauge boson $V$. The process on the right panel can only arise for $S$ coupling directly to the light quarks. It has not been considered in our analysis, since it can be neglected under simple flavor assumptions (see the text for details).}
\label{fig:diagass}
\end{figure}

\begin{figure}[!ht]
\hspace{-0.2cm}\centerline{\includegraphics[width=\columnwidth]{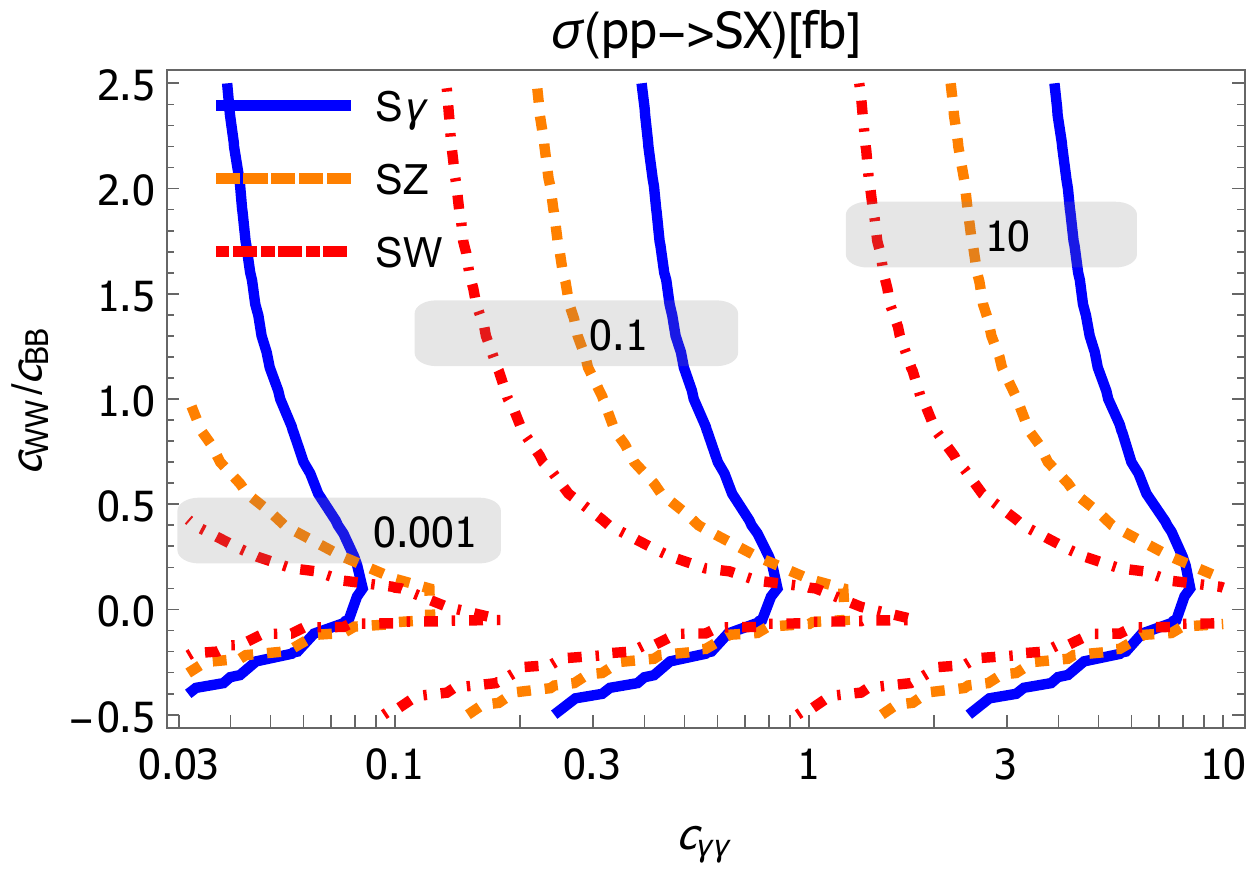}}
\caption{Contour lines of the cross sections (in\,fb) for the associated production channel with an electroweak gauge boson in the plane $c_{\gamma\gamma}-c_{WW}/c_{BB}$. The numbers stand for each group of three contour lines.}
\label{fig:assxsec}
\end{figure}

\begin{figure}[!ht]
\hspace{-0.35cm}\centerline{\includegraphics[width=\columnwidth]{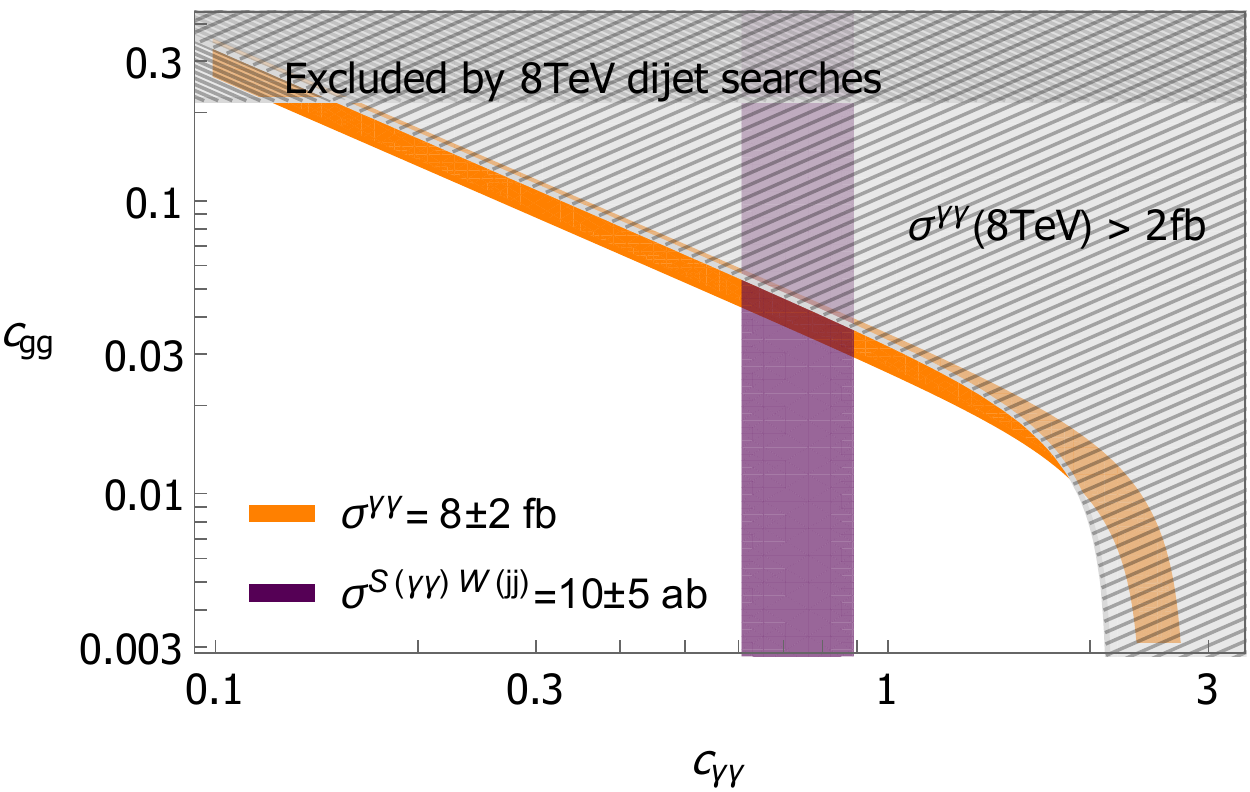}}
\caption{Regions in the $c_{\gamma\gamma}-c_{gg}$ plane constrained by the diphoton excess for $c_{WW}/c_{BB} = 1$ for $\Gamma_S = 45$\,GeV.  The vertical band stands for  the value of $c_{\gamma\gamma}$ determined by measurements of $p p \rightarrow SW^\pm \rightarrow 2\gamma\,2j$ (see the text for details). Dijet constraints from 8\,TeV data~\cite{Aad:2014aqa,CMS:2015neg} are also shown.}
\label{fig:flatdir}
\end{figure}

It can be seen from Fig.~\ref{fig:parameterspace} that resonance searches for massive gauge bosons are only sensitive to the ratio $c_{WW}/c_{BB}$. It is also apparent that there are other flat directions, i.e. that the different couplings cannot be accessed independently from each others. In fact, even if it was possible to determine $\Gamma_S$ experimentally, we would have to measure all $S$ decay modes to be able to bound each coupling independently. This seems highly unrealistic, first, because $\Gamma_S$ might remain out of the experimental resolution, and second because  it would require to also tag decays into gluons and (potentially) invisible particles, a notoriously difficult task in the busy hadronic environment of the LHC. Thus, different strategies should be considered in this respect. One possibility relies on $S$ production in association with a gauge boson (a previous study in this direction has been presented in~\cite{Alves:2015jgx}). The corresponding Feynman diagram is depicted in the left panel of Fig.~\ref{fig:diagass}, while the cross sections are shown in Fig.~\ref{fig:assxsec}. These have been computed by using \texttt{MadGraph v5}~\cite{Alwall:2014hca} (\texttt{Feynrules v2}~\cite{Alloul:2013bka} has been first used to implement the interactions of Eq.~\ref{eq:par}). Other automatic tools have been developed and used~\cite{Staub:2016dxq} to study models aiming at explaining the diphoton excess. In the region of parameter space compatible with the reported excess, the associated production cross sections can be as large as few tens fb. And even for rare decay modes, (e.g.  a branching ratio below 0.001 for $S\rightarrow ZZ \rightarrow 4\ell$), enough events can still be collected with large luminosities. Note also that the corresponding backgrounds are almost negligible (see for example~\cite{Aad:2015bua} for an experimental study of three photon final states). Thus, in Fig.~\ref{fig:flatdir}, we elaborate on the idea of resolving flat directions using further production modes. To this end, we consider a hypothetical scenario in which the ratio $c_{WW}/c_{BB}$ has been experimentally established (this measurement can be performed by just observing the ratio of $\gamma\gamma$ events over $ZZ$ or $Z\gamma$ events). Clearly, $c_{\gamma\gamma}$ and $c_{gg}$ cannot just be individually determined by fitting the diphoton excess. This flat direction in the $c_{gg}-c_{\gamma\gamma}$ plane is depicted by the orange band in Fig.~\ref{fig:flatdir} for $\Gamma_S = 45$\,GeV and $c_{WW}/c_{BB} = 1$. Now in addition if the associated production $SW^\pm \rightarrow 2\gamma\, 2j$ is observed to be, for example, $0.01\pm 0.005$\,fb, the degeneracy is broken and $c_{\gamma\gamma}$ can be constrained independently, as shown by the vertical band in the figure. 
The discussion above assumes no substantial direct coupling of $S$ to the SM fermions. If such couplings exist, 
another contribution to the associated production originates  when an EW gauge boson is radiated off from one of the initial quarks (right panel of Fig.~\ref{fig:diagass}). However, the cross section is negligible when linked to light quarks under the assumption that   the couplings obey a minimal flavor violation structure and are therefore naturally expected to be of the size of the Yukawa couplings. Flavor constraints would be hard to evade otherwise.  If large couplings to the light quarks were nonetheless present, relying on some cancellation to pass flavor constraints, then a detailed study of the kinematic  would be worth performing in order to discriminate the various contributions to the associated production. We have checked that the associated production cross section via an initial $b$ quark remains smaller than the contributions computed in Fig.~\ref{fig:assxsec} in most of the parameter space. Finally, we have checked that the gluon-fusion associated production $S+W^\pm/Z/\gamma$ together with an extra jet is typically subdominant too, except in the region of small $c_{\gamma\gamma}$ ($<0.1$) where it can anyway be reduced by an appropriate cut on the gauge boson $p_T$ and by vetoing the extra jet.

This simple analysis illustrates the importance of considering the associated production mechanisms. Indeed, the argument does not hold equally well for $S$ production in VBF since it turns out to have a remaining large dependence on $c_{gg}$. The reason is that, contrary to the Higgs case whose couplings to the electroweak gauge bosons appear at the tree level, VBF contamination by gluon initiated processes in the singlet case can be rather large even after tagging on forward jets~\cite{Bjorken:1992er, DelDuca:2006hk}. Cuts in this respect are provided in section~\ref{sec:vbf}. Nonetheless, it is worth to point out that measurements in VBF together with the determination of $\Gamma_S$ might shed light on possible $S$ hidden decays~\cite{Kamenik:2016tuv}.

On top of it, a last comment concerns the spin-1 alternatives for explaining the diphoton excess. As it has been pointed out in \cite{Chala:2015cev}, these scenarios rely on the production of a 750\,GeV vector boson that subsequently decays into a photon and a light scalar. The latter further decays into two collimated photons that, at the detector level, appear to be a single one. This kind of setup can not however give rise to sizable amount of three gauge boson events. Particularly with $W^\pm$ in the final state. As a consequence, $S$ production in association with gauge bosons provides a striking signature for disentangling spin-0 and spin-1 models.

The distinctive kinematics of associated production provides different ways to inquire the parity of such a scalar. The polar angle of the radiated vector boson has been highlighted in this respect in the context of Higgs physics (see for example~\cite{Miller:2001bi,Choi:2002jk}, and ~\cite{Aaltonen:2015mka} for related experimental searches at Tevatron). The large Higgs coupling to the longitudinal polarization of the gauge bosons are however instrumental for these studies. The rather small splitting between the Higgs mass and $m_Z$ and $m_W$ is also of major significance for analyses based on the behavior of the cross section near threshold. Accordingly, this observable is no longer suitable for $S$ physics (note that $S$ might not even couple to the longitudinal polarization of the gauge bosons). Related results in this direction have been also pointed out in~\cite{Djouadi:2016eyy}.
Further observables for Higgs physics have been presented for example in~\cite{Godbole:2013lna}. Several angles between the Higgs momentum and  reconstructed momenta of the leptons and jets in the decay of the gauge boson produced in association with the Higgs boson have been identified as discriminating variables to scrutinize the CP properties of $S$. However, the rather small cross section in this channel compared to GF and VBF 
makes the latter much more appropriate for an early data analysis. We will thus focus on these channels hereafter. 

\section{Gluon fusion}
\label{sec:gf}
The GF production cross section can be conveniently written as
\begin{equation}\label{eq:gfx}
 \sigma^\text{GF} = 123\times\left(\frac{c_{gg}}{0.01}\right)^2\,\text{fb},
\end{equation}
as computed at LO using \texttt{MadGraph}. The \texttt{NN23LO1}~\cite{Ball:2012cx} parton-distribution functions (PDFs) have been used. From the computation of the GF production at higher order in the SM, we expect a large K-factor of order $1.7-2$ at NLO. This K-factor will anyway drop in the computation of the asymmetry computed below. We do not include it since a consistent treatment would also require a NLO estimation of the various backgrounds, which is beyond the scope of our analysis.
Three different decay modes of $S$ are considered in GF, namely $S\rightarrow ZZ$ in both the fully leptonic ($4\ell$) and the semileptonic channels ($2j\, 2\ell$) as well as $S\rightarrow W^+W^-$ with semileptonic decay ($2j\ell\,\slashed{E}_T$). In order to tag these events at the experimental level, all events are first required to pass the following set of common cuts. Leptons must have $p_T^\ell > 10$\,GeV and $|\eta_\ell| < 2.5$. Jets are instead required to have $p_T^j > 20$\,GeV and $\eta_j < 5$. Same-flavor leptons must be separated by $\Delta R > 0.2$ while different-flavor leptons must fulfill $\Delta R > 0.1$. Besides, all leptons must be separated from other jets by $\Delta R > 0.2$, and jets by $\Delta R > 0.4$ among themselves.

Then, exactly two opposite-sign lepton pairs are required in the four-lepton channel. The two with invariant mass closest to $m_Z$ are tagged as coming from one $Z$, and the other two from the second one. In the semileptonic $ZZ$ decay exactly two opposite-sign leptons and at least two jets must be present. In the semileptonic $W^+ W^-$ decay exactly one lepton and at least two jets and $\slashed{E}_T > 20$\,GeV are instead required. The longitudinal momentum of the neutrino can be in this case reconstructed by the $W^\pm$ on-shell condition (see for example~\cite{daSilvaFernandesdeCastro:2008zz}). Following Ref.~\cite{Khachatryan:2015cwa}, we take the smaller in absolute value among the two possible solutions. There is no further ambiguity in pairing pair of particles with the corresponding massive gauge bosons in any of these channels. We therefore require any reconstructed $Z$ ($W^\pm$) mass to be in a window of $\pm 20$\,GeV around $m_Z$ $(m_W)$. Besides, each $Z$ and $W^\pm$ is required to have $p_T > 250$\,GeV.
\begin{table}[t]
\hspace{0.2cm}\begin{tabularx}{0.95\columnwidth}{XXXX}
 & $4\ell$ & $ 2j\,2\ell $ & $2j\,\ell\, \slashed{E}_T$\\[0.1cm]\hline\\[-0.2cm]
$\epsilon$ (\%)  & 42 & 40 & 30 \\[0.1cm]
$\sigma_b$ (fb)  & 0.04 & 34 & 240 \\[0.1cm]\hline
\end{tabularx}
\caption{Estimated signal efficiencies ($\epsilon$) and background cross sections ($\sigma_b$) for GF events after the cuts described in the text.}
\label{tab:effgf}
\end{table}
On top of this, the invariant mass of the four SM tagged particles is required to be in the range $[700, 800]$\,GeV. Finally, the event must not pass the VBF criteria, to be defined in the next section.

The window around the mass of the gauge bosons in the previous cuts is required in order to further reduce the background with respect to the signal (which gets only slightly affected). In particular, pair of
jets come mainly from QCD radiation and hence their invariant masses do not necessarily peak at the $m_W (m_Z)$ mass. However, if signals with of-shell gauge bosons were to be studied, the corresponding cut
should be relaxed. In this respect, it is worth mentioning Ref.~\cite{Stolarski:2016dpa}, where the authors claim that the four-lepton channel can get sizable contributions from processes containing a virtual photon.

The efficiencies for selecting events in each of these categories are shown in Table~\ref{tab:effgf}. The estimated cross sections for the SM backgrounds after passing all cuts are also shown. The irreducible backgrounds dominate in all cases. Therefore, only these have been taking into account. In order to compute all these quantities we have generated parton-level events with \texttt{MadGraph v5} which are subsequently passed through \texttt{Pythia v6}~\cite{Sjostrand:2006za} to account for showering, hadronization and fragmentation effects. The cuts above are finally implemented in \texttt{MadAnalysis v5}~\cite{Conte:2014zja}.
\begin{figure}[!ht]
\centerline{ \includegraphics[width=\columnwidth]{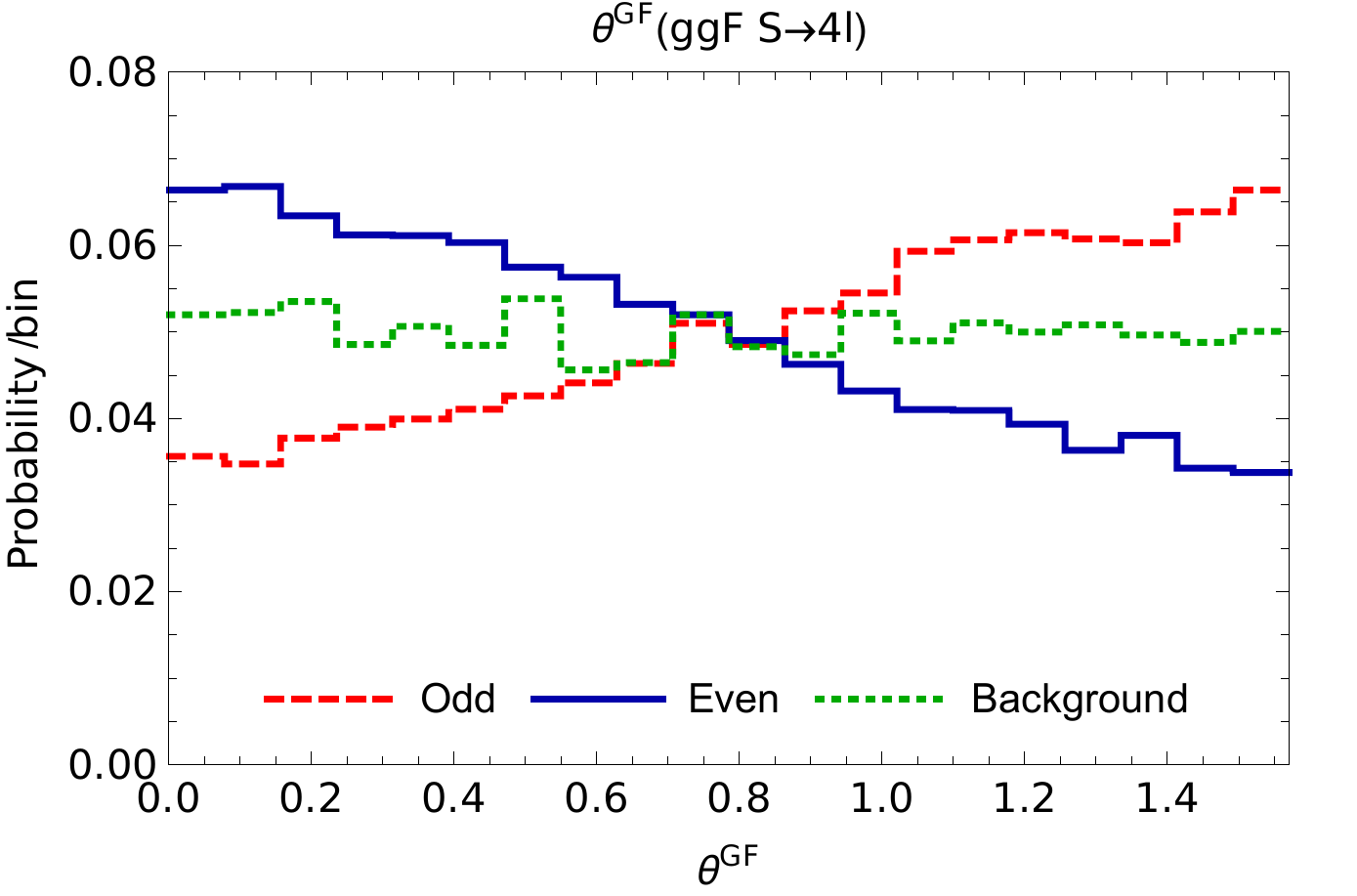}}
 \caption{$\theta^\text{GF}$ distribution for reconstructed four-lepton signal events for the scalar (solid blue) and the pseudo-scalar (dashed red) cases. The background is shown in dotted green. Signal and background distributions have been independently normalized to unity. Their respective importance will depend on the parameter space point.}\label{fig:dgf}
\end{figure}
\begin{figure}[!ht]
 \includegraphics[width=\columnwidth]{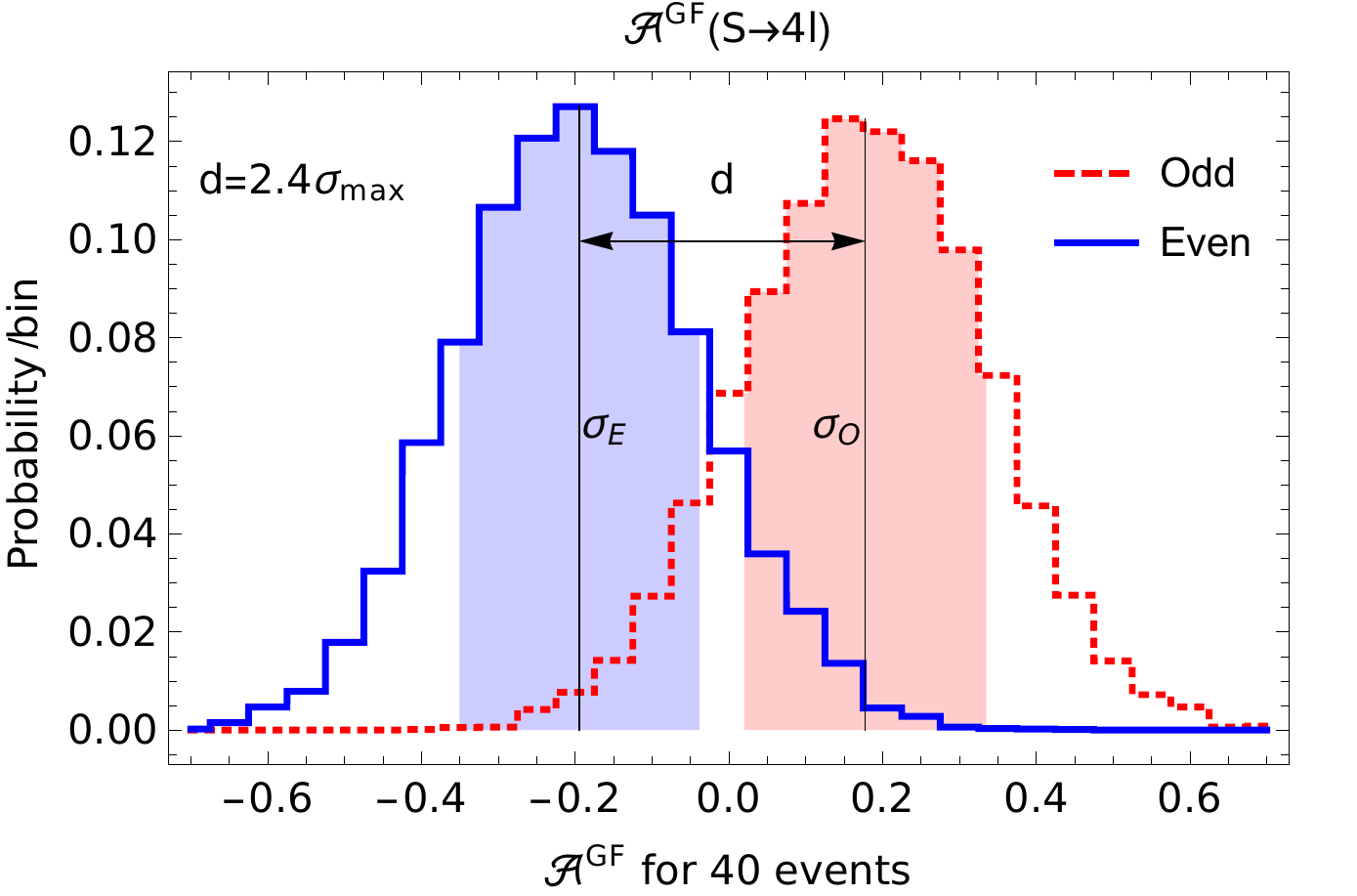}
 \caption{$\mathcal{A}^\text{GF}$ distribution for the scalar (solid blue) and the pseudo-scalar (dashed red) cases with 40 observed events after 10'000 pseudo experiments. The distance ($d$) between the two central values in terms of the largest $\sigma$ is also shown.}\label{fig:dagf}
\end{figure}

Having reconstructed the momenta of the four decay products, we can define the following asymmetry:
\begin{equation}\label{eq:gf}
 \mathcal{A}^\text{GF} = \frac{N(\theta^\text{GF} > \pi/4) - N(\theta^\text{GF} < \pi/4)}{N(\theta^\text{GF} > \pi/4) + N(\theta^\text{GF} < \pi/4)},
\end{equation}
where
\begin{equation}\label{eq:thetagf}
 \theta^\text{GF} = \bigg\lbrace\begin{array}{llllll}
                                 \theta & & \text{if} & & & \theta < \pi/2\\
                                 \pi - \theta &  & \text{if} & & & \theta > \pi/2
                                \end{array},
\end{equation}
and
\begin{equation}
 \theta = \arccos{\left\lbrace\frac{(p_1\times p_2)\cdot (p_3\times p_4)}{|p_1\times p_2| |p_3\times p_4|}\right\rbrace},
\end{equation}
with $p_{1,2}$ and $p_{3,4}$ the three-momenta of the decay products of each massive gauge boson. This observable has been widely used in Higgs physics (see for example~\cite{ATLAS:2015zja}). However, the small Higgs mass makes some channels above not suitable for CP studies with this asymmetry,  inasmuch as the signal peaks in the region populated by the SM background. For $S$ decays instead, the rather large mass allows us to stay in much more suppressed background regions. Note also that two body $S$ decays could be also considered. As a matter of fact,  photon conversion events have been discussed in the Higgs literature~\cite{Bishara:2013vya}. The typical opening angle of the lepton products are however of the order of $m_e/E_\gamma\sim 10^{-6}$, which is well below any present or future experimental sensitivity.
\begin{figure*}[!ht]
\hspace{-0.4cm}\centerline{\includegraphics[width=1.1\columnwidth]{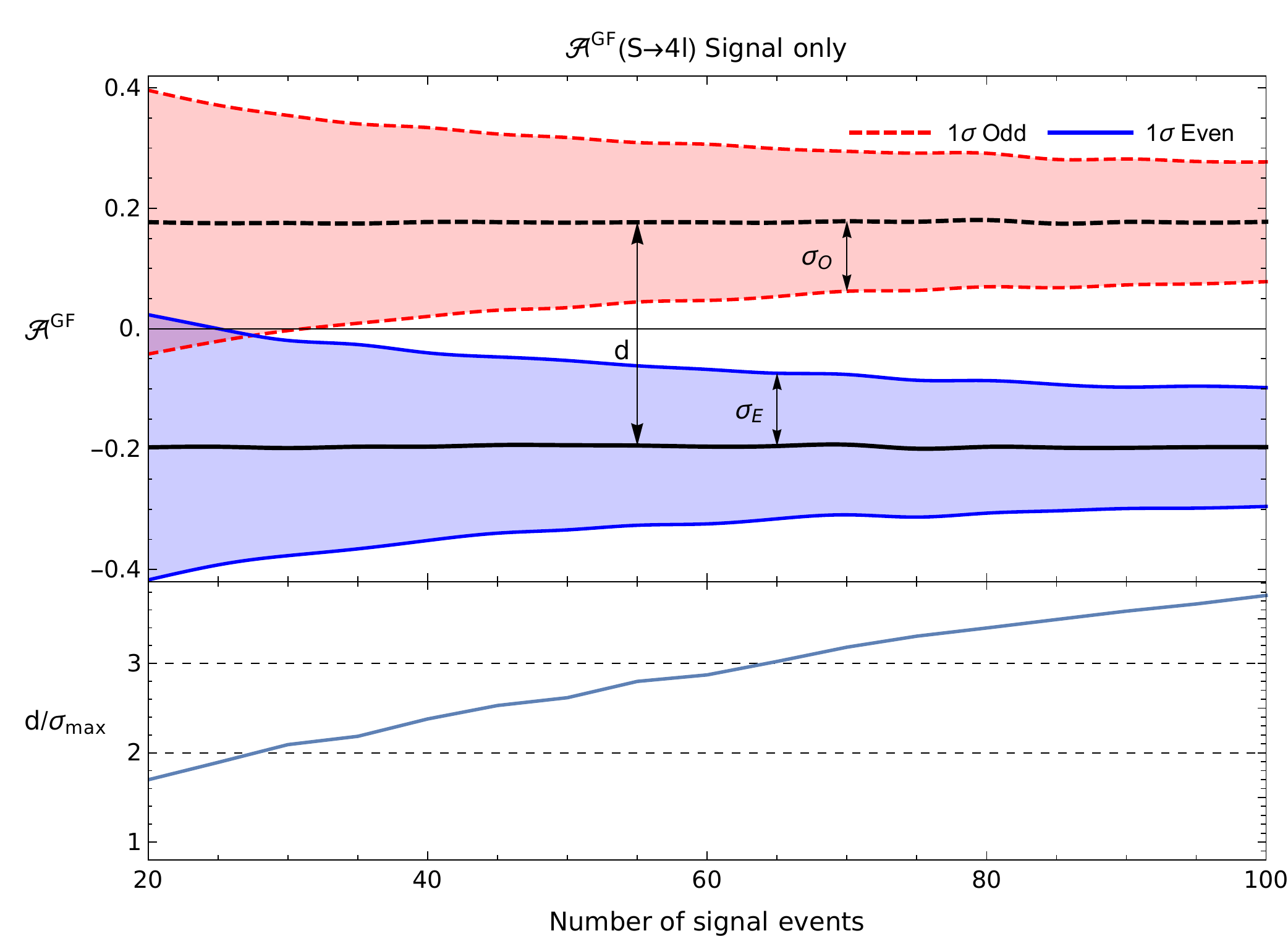}\hspace{0.cm}
  \includegraphics[width=1.1\columnwidth]{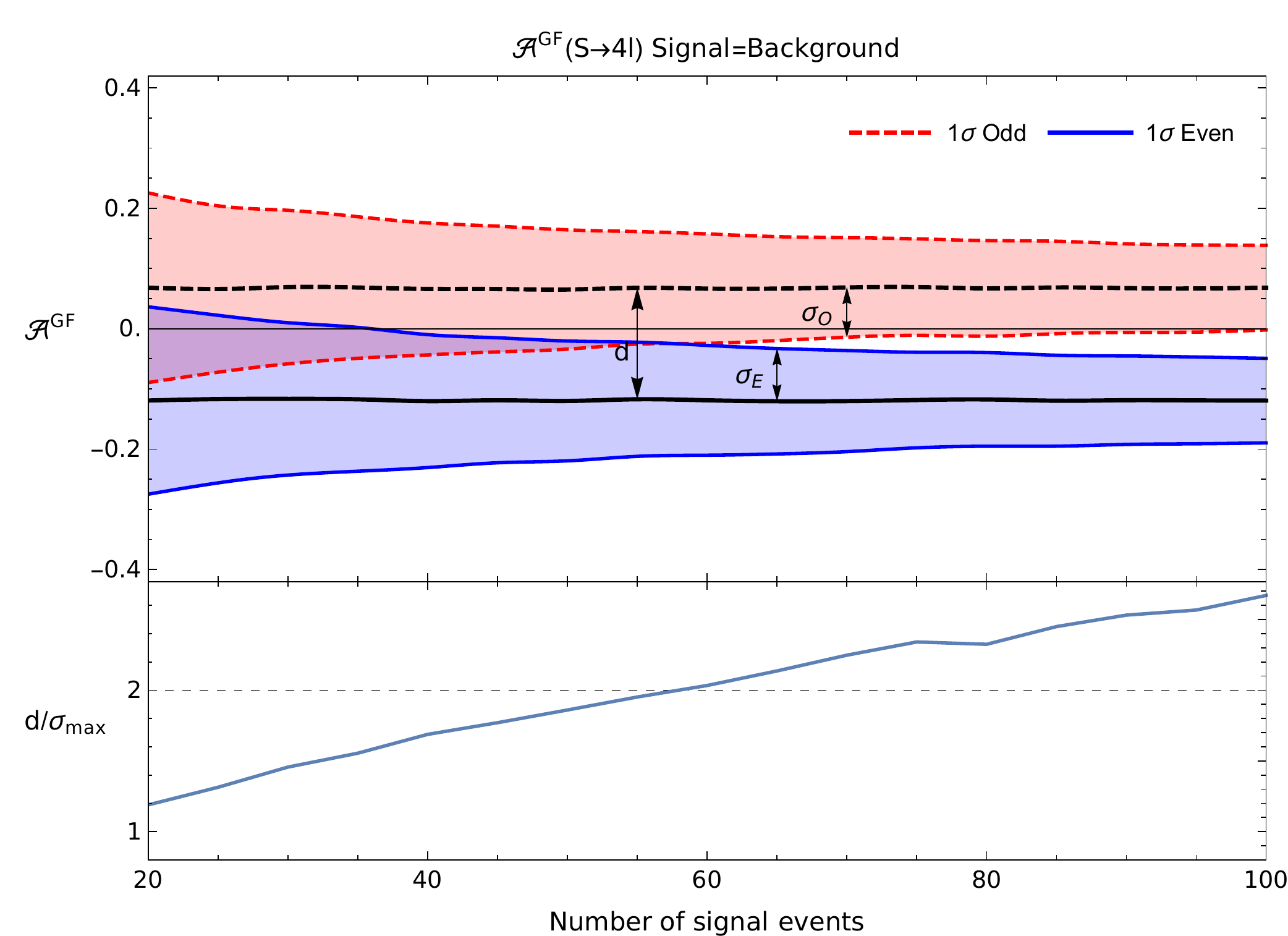}}
 \caption{One sigma statistical interval for $\mathcal{A}^\text{GF}$  as a function of the total number of observed events for only signal (left panel) and with as much background as signal (right panel) for the scalar (solid blue) and pseudo-scalar (dashed red) cases. The distance ($d$) between the two central values in terms of $\sigma_{\text{max}} \equiv \text{max}\lbrace\sigma_O, \sigma_E\rbrace$ is also shown.}\label{fig:resgf}
\end{figure*}

In four-lepton events, the variable defined in Eq.~\ref{eq:thetagf} takes the form shown in Fig.~\ref{fig:dgf}. No significant departures from this shape are found in other channels.
In order to quantify the discrimination power of this asymmetry for a given number $N_\text{obs}$ of observed events, we perform 10'000 pseudo experiments with $N_\text{obs}$ events each. As a matter of example, the distribution followed by $\mathcal{A}^\text{GF}$ for $N_\text{obs} = 40$ for signal only is shown in Fig.~\ref{fig:dagf}. The one  sigma statistical uncertainty is defined by the symmetric interval around the center of the distribution containing the 68\% of the total area. For the matter of example, this is also shown in the figure. These 40 events in the  $2j\ell\slashed{E}_T$ final state can be reached, in the minimal width case, for luminosities as low as $\mathcal{L} \sim 5$\,fb$^{-1}$ for $(c_{\gamma\gamma},c_{WW}/c_{BB}) = (2,2)$, while $(c_{\gamma\gamma},c_{WW}/c_{BB}) = (0.1,0.1)$ would require $\mathcal{L} > 100$\,fb$^{-1}$. The large background makes the analysis harder and the luminosity needed to discriminate the two CP hypotheses will be estimated in section~\ref{sec:results}. Figure~\ref{fig:resgf} gives the asymmetry $\mathcal{A}^\text{GF}$ as a function of the total number of observed events under the assumption of negligeable background (left panel) and under the assumption of as many background events as signal events (right panel).

%
%
\section{Vector-boson fusion}
\label{sec:vbf}
\begin{figure}[t]
\hspace{0.3cm}\centerline{\includegraphics[width=1.1\columnwidth]{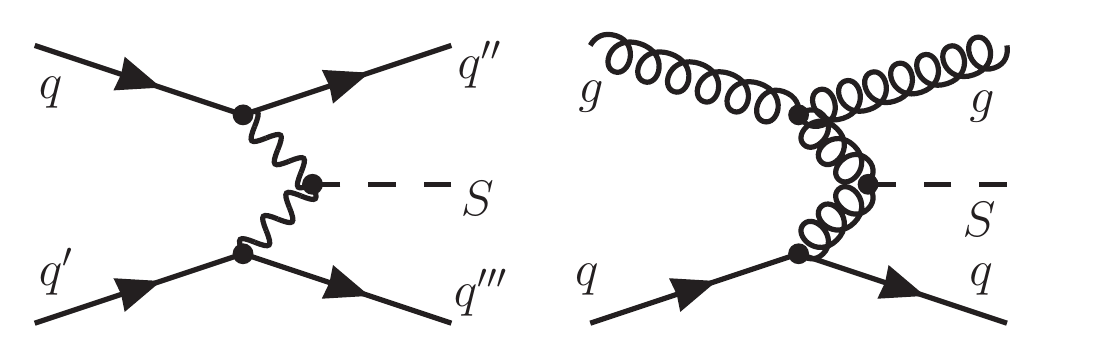}}
\caption{Examples of Feynman diagrams for $S$ production via electroweak (left) and QCD (right) with two additional jets.}
\label{fig:diagvbf}
\end{figure}

\begin{figure}[!ht]
\hspace{-0.2cm}\centerline{\includegraphics[width=\columnwidth]{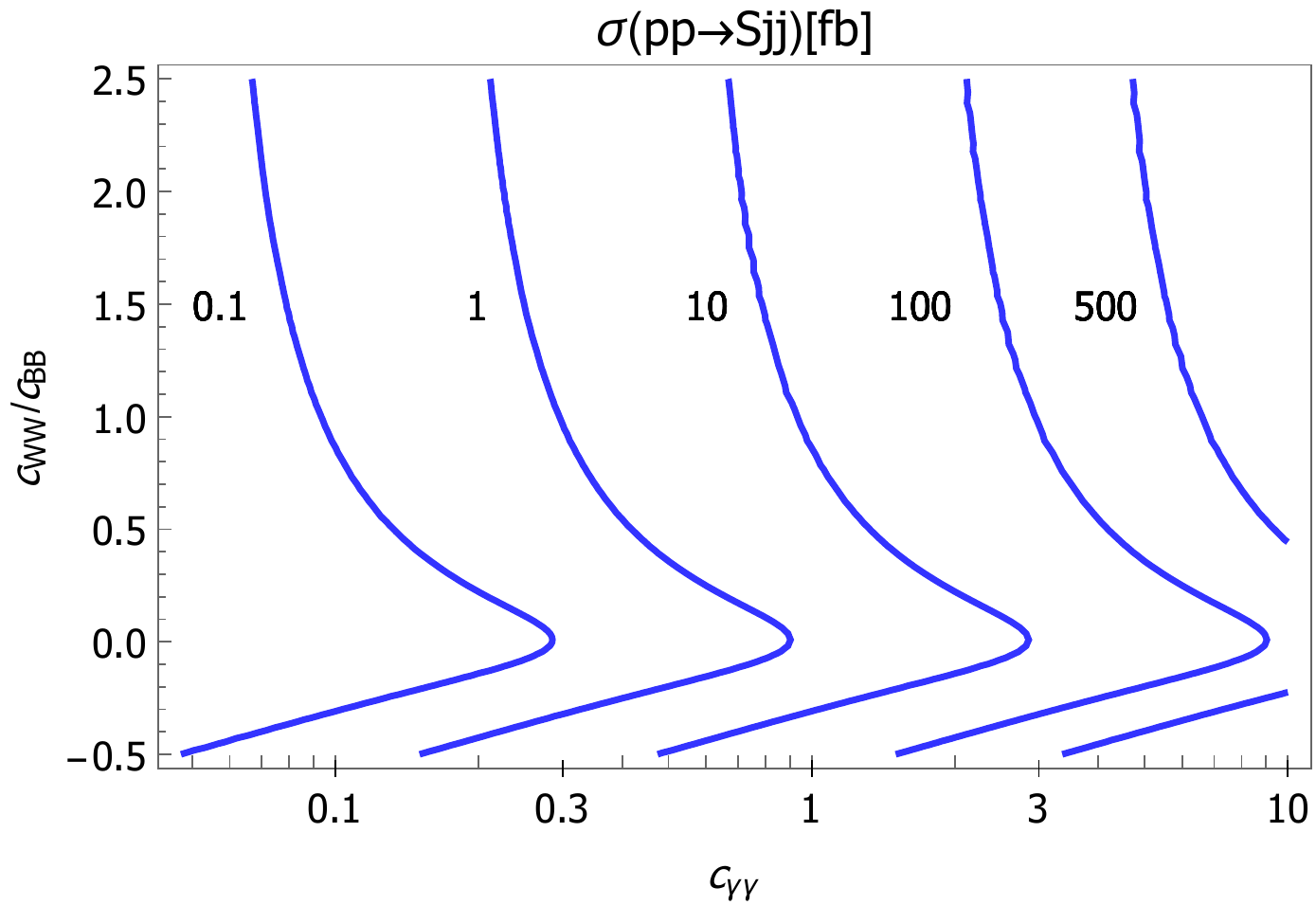}}
\caption{Contour lines in the plane $c_{WW}/c_{BB}-c_{\gamma\gamma}$ of $S$ production cross sections (in\,fb) with two forward jets initiated by electroweak gauge bosons and after the parton-level VBF cuts described in the text.}
\label{fig:vbfxsecs}
\end{figure}

The LO cross section for producing $S$ in association with two jets with $p_T$ larger than 10\,GeV, separated by at least $\Delta R > 0.1$ and with dijet invariant mass above 400\,GeV, can be approximately written as
\begin{align}\label{eq:vbfeq}\nonumber
 \sigma^{\text{VBF}} = \,& \bigg[45~ \left(\frac{c_{gg}}{0.01}\right)^2 + 1.2 \frac{c_{\gamma\gamma}^2}{(1 + r)^2} \\
 &+ 1.7 \frac{c_{\gamma\gamma}^2 r}{(1 + r)^2} + 43 \frac{c_{\gamma\gamma}^2 r^2}{(1 + r)^2}\bigg]\,\text{fb},
\end{align}
with $r \equiv c_{WW}/c_{BB}$. The coefficients above have been again computed using \texttt{MadGraph} with the \texttt{NN23LO1} PDFs. The interference between gluon-initiated diagrams (proportional to $c_{gg}$) and VBF diagrams is negligible and hence not shown in this equation. Two example diagrams are depicted in Fig.~\ref{fig:diagvbf}. Hereafter we denote by $S^{\text{QCD}}$ and $S^{\text{EW}}$ the production computed using each channel alone. $S^{\text{QCD}}$ in the plane $c_{\gamma\gamma}-c_{WW}/c_{BB}$ can be easily estimated using this equation in light of the $c_{gg}$ values provided in Fig.~\ref{fig:parameterspace}. Instead $S^{\text{EW}}$ is  plotted in Fig.~\ref{fig:vbfxsecs}.

VBF events can be tagged at the experimental level in five different decay modes of $S$. These comprise the three possibilities described in the previous section with two additional forward jets, namely $4\ell\, 2j$, $4j\, 2\ell$ and $4j\,\ell\, \slashed{E}_T$, as well as the decay into $\gamma\gamma$ and $\ell^+\ell^-\,\gamma$.
Events are first selected by imposing the same common cuts as in GF, while photons should be separated from any other tagged particle by $\Delta R > 0.2$. When more than two jets are present, forward-jet candidates are selected to be those two jets with invariant mass $m_{j_1j_2}$ less similar to $m_Z$ (or $m_W$) among the four leading jets. They are subsequently required to fulfill the VBF criteria. This is defined by $m_{j_1j_2} > 500$\,GeV, $\eta_{j_1}\eta_{j_2} < 0$, $|\Delta\eta_{j_1 j_2}| > 3$ and $\Delta R_{j_1j_2} > 0.4$. These cuts are motivated by previous searches for heavy Higgs bosons~\cite{Aad:2015kna}. Any reconstructed $Z$ ($W^\pm$) is again required to have a mass within a window of $\pm 20$\,GeV
around $m_Z$ ($m_W$). Besides, the $p_T$ of the two leading photons as well as the $p_T$ of each reconstructed $Z$ and $W^\pm$ must be still larger than 250\,GeV. Finally, we require the invariant mass of the two reconstructed SM gauge bosons to be in the range [700, 800]\,GeV. The efficiencies for selecting events in each of these categories, referred to events generated using the parton-level cuts in Eq.~\ref{eq:vbfeq}, are shown in Table~\ref{tab:effvbf} for $S^{\text{QCD}}$ and $S^{\text{EW}}$. Notice that gluon-initiated VBF can be dominant due to the rather large coefficient in front of $c_{gg}$ in Eq.~\ref{eq:vbfeq}. This result contrasts with the Higgs case, the reason being that, unlike the singlet $S$, the Higgs boson couples to the electroweak gauge bosons at the tree level.

\begin{table}[!ht]
\begin{tabularx}{1.0\columnwidth}{XXXXXX}
 & $2\gamma\, 2j$ & $4\ell\, 2j$ & $2\ell\,\gamma\, 2j$  & $ 4j\,2\ell $ & $4j\,\ell\,  \slashed{E}_T$\\[0.1cm]\hline\\[-0.2cm]
$\epsilon_{\text{QCD}}$  & 6 & 12 & 14 & 11 & 9\\[0.1cm]
$\epsilon_{\text{EW}}$  & 18 & 15 & 18 & 15 & 12\\[0.1cm]
$\sigma_b\,\,(\text{fb})$  & 0.42 & 0.001 & 0.03 & 1.8 & 15\\[0.1cm]\hline
\end{tabularx}
\caption{Estimated signal efficiencies ($\epsilon$, in percent) and background cross sections ($\sigma_b$) for VBF events after the cuts described in the text. Gluon-initiated processes (QCD) can very much contaminate pure electroweak (EW) VBF.}
\label{tab:effvbf}
\end{table}
The estimated background cross sections are also shown in Table~\ref{tab:effvbf}. The irreducible backgrounds dominate each category. The channels $4j\, 2\ell$ and $4j\,\ell\,\slashed{E}_T$ are mostly populated by Drell--Yan and $W^\pm$ production with radiated jets, rather than by diboson events. These cross sections are of similar magnitude to those reported in the figures of Ref.~\cite{Aad:2015kna}, which uses slightly different cuts.
We construct the following asymmetry for VBF events:
\begin{equation}
 \mathcal{A}^\text{VBF} = \frac{N(\theta^\text{VBF} > \pi/4) - N(\theta^\text{VBF} < \pi/4)}{N(\theta^\text{VBF} > \pi/4) + N(\theta^\text{VBF} < \pi/4)},
\end{equation}
where, analogously to the GF case,
\begin{equation}
 \theta^\text{VBF} = \bigg\lbrace\begin{array}{llllll}
                                 \theta & & \text{if} & & & \theta < \pi/2\\
                                 \pi - \theta &  & \text{if} & & & \theta > \pi/2
                                \end{array},
\end{equation}
\begin{figure}[!ht]
\centerline{\includegraphics[width=\columnwidth]{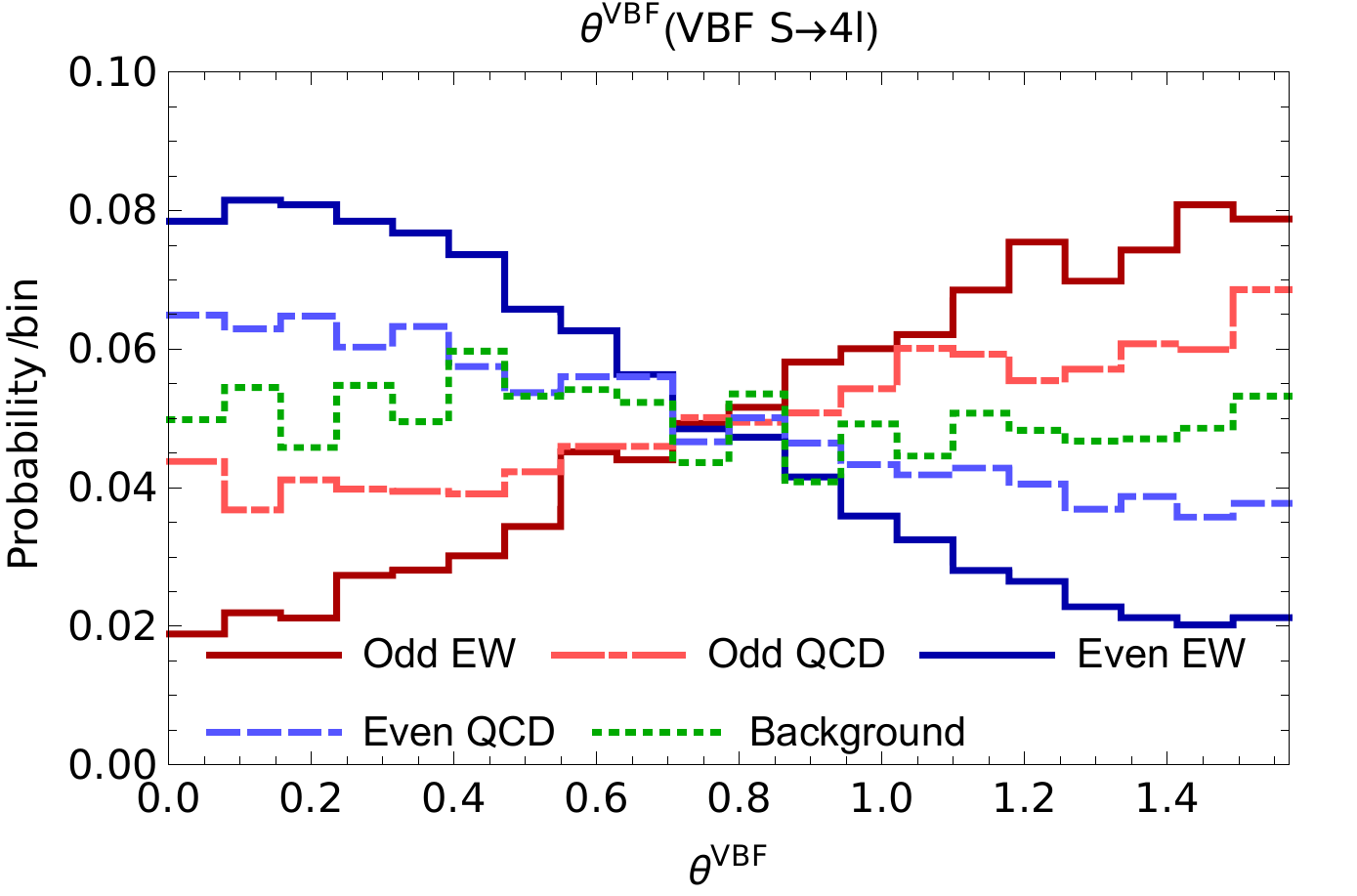}}
 \caption{$\theta^\text{VBF}$ distribution for reconstructed four-lepton signal events for the scalar (blue) and the pseudo-scalar (red) cases. Solid and dashed lines stand for $S^\text{EW}$ and $S^\text{QCD}$ respectively. The background is shown in dotted green. Signal and background distributions have been independently normalized to unity. Their respective importance will depend on the parameter space point.}\label{fig:dvbf}
\end{figure}
\begin{figure}[!ht]
 \includegraphics[width=\columnwidth]{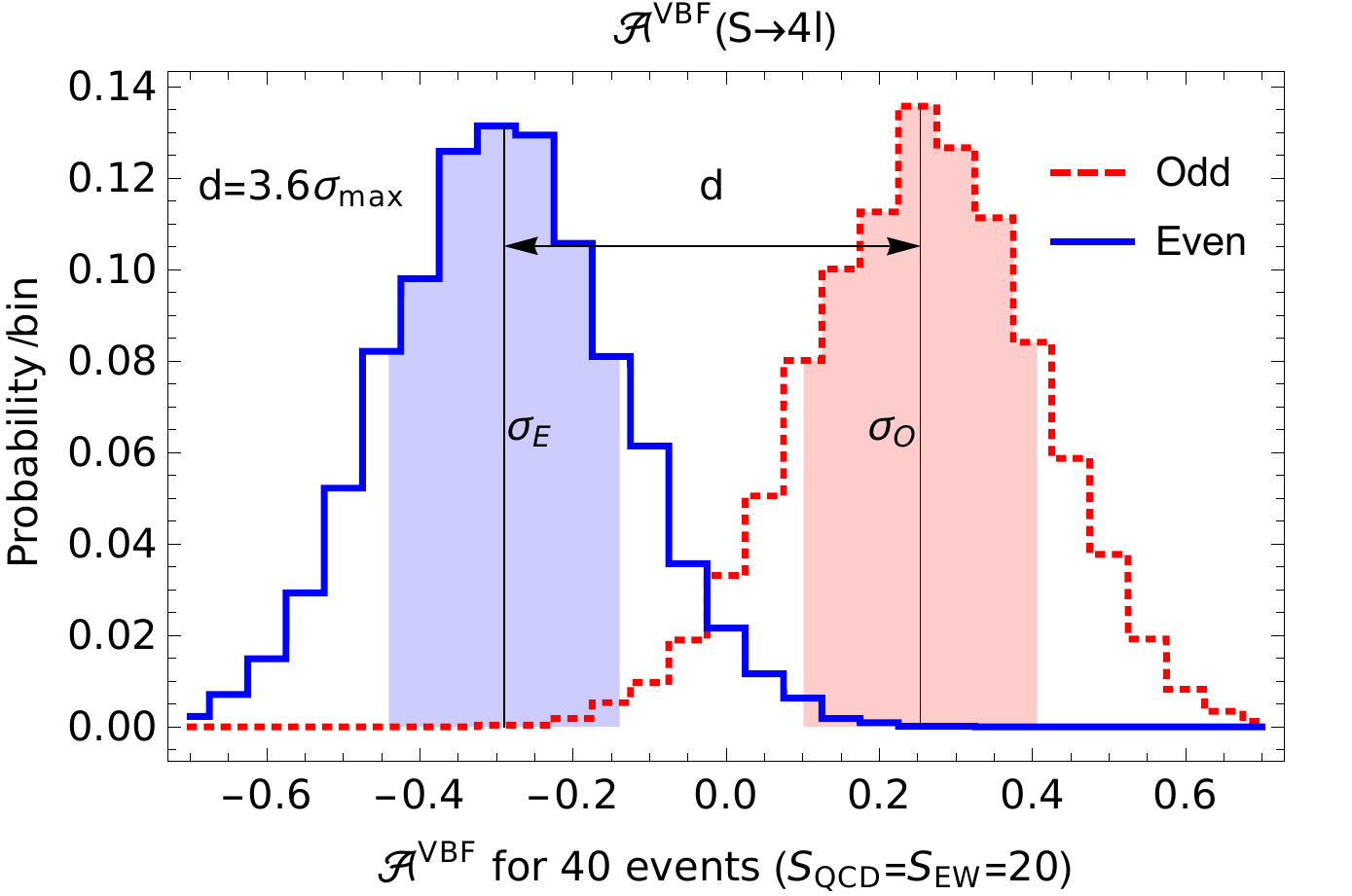}
 \caption{$\mathcal{A}^\text{VBF}$ distribution for the scalar (solid blue) and the pseudo-scalar (dashed red) cases with 40 observed events and $S^\text{QCD} = S^\text{EW}$ after 10'000 pseudo experiments. The distance ($d$) between the two central values in terms of the largest $\sigma$ is  shown in the lower panels.}\label{fig:davbf}
\end{figure}
\begin{figure*}[t]
\hspace{-0.5cm}\centerline{\includegraphics[width=1.1\columnwidth]{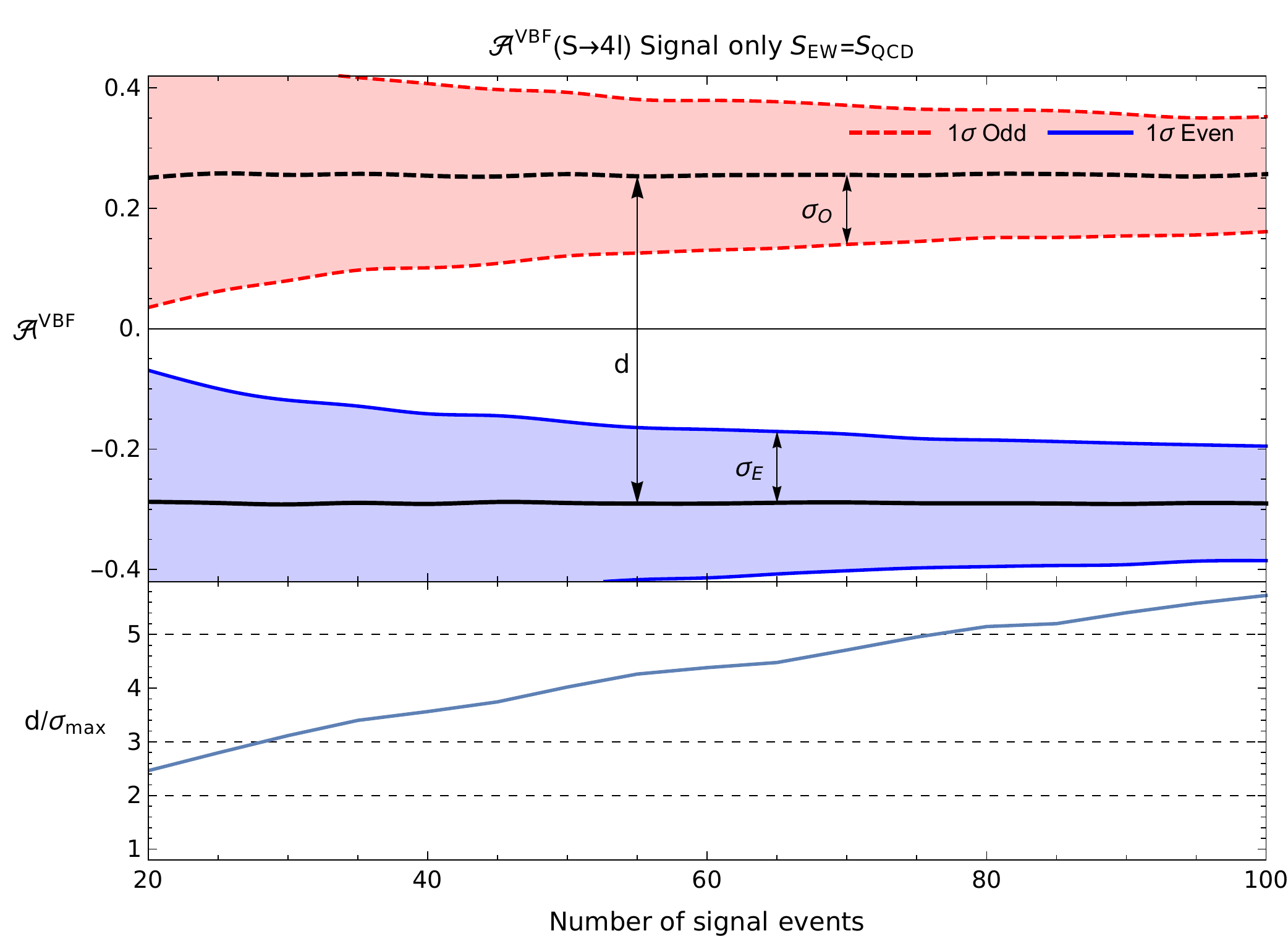}\hspace{0.cm}
  \includegraphics[width=1.1\columnwidth]{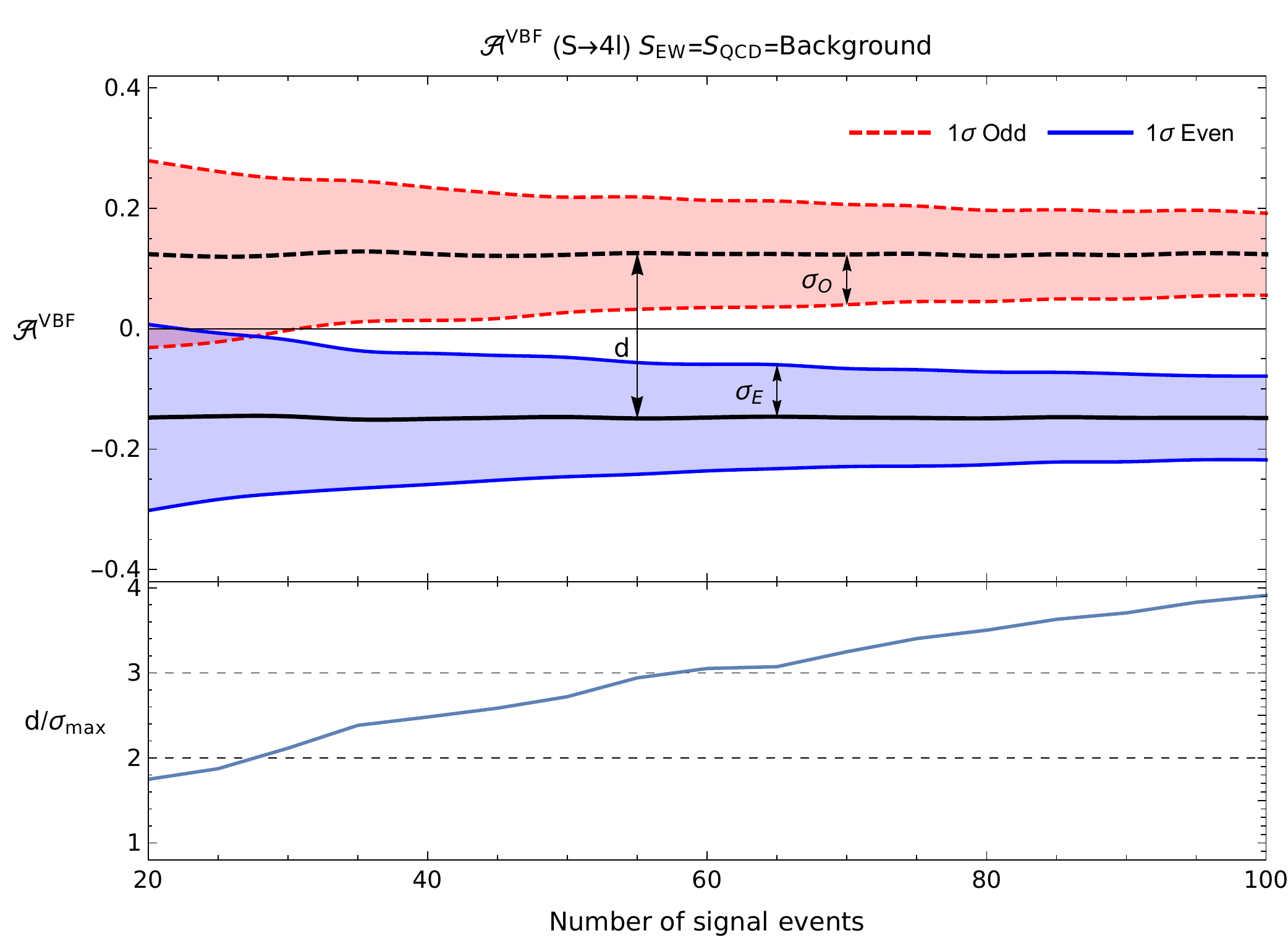}}
 \caption{One sigma statistical interval for $\mathcal{A}^\text{VBF}$ as a function of the total number of observed events for only signal (left panel) and as much background as signal (right panel) for the scalar (solid blue) and pseudo-scalar (dashed red) cases. $S^\text{QCD} = S^\text{EW}$ has been assumed in both panels. The distance ($d$) between the two central values in terms of  $\sigma_{\text{max}} \equiv \text{max}\lbrace\sigma_O, \sigma_E\rbrace$ is shown in the lower panels.}\label{fig:resvbf}
\end{figure*}
\begin{figure*}[t]
\hspace{-0.2cm}\centerline{\includegraphics[width=1.02\columnwidth]{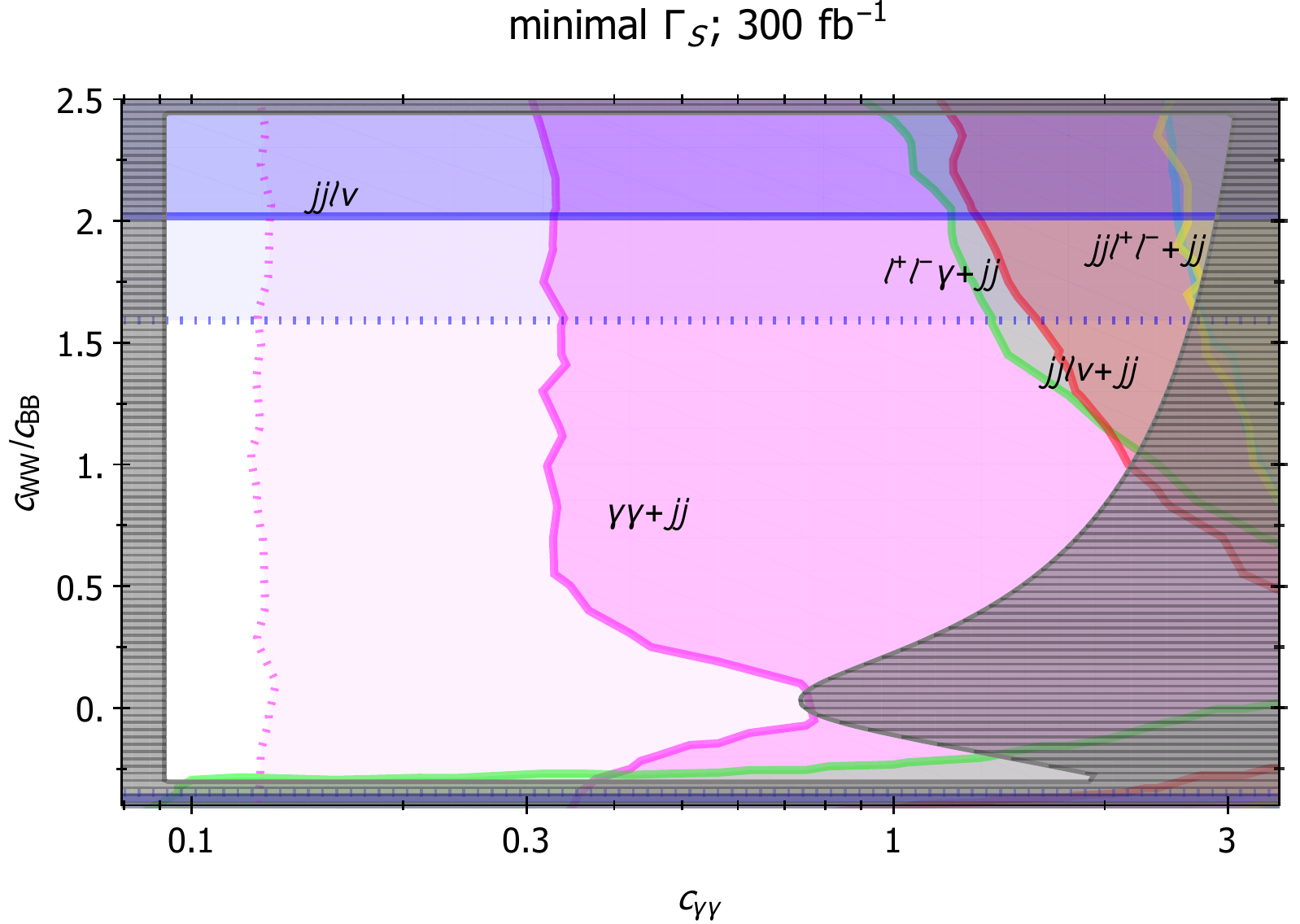}\hspace{0.5cm}
 \includegraphics[width=1.02\columnwidth]{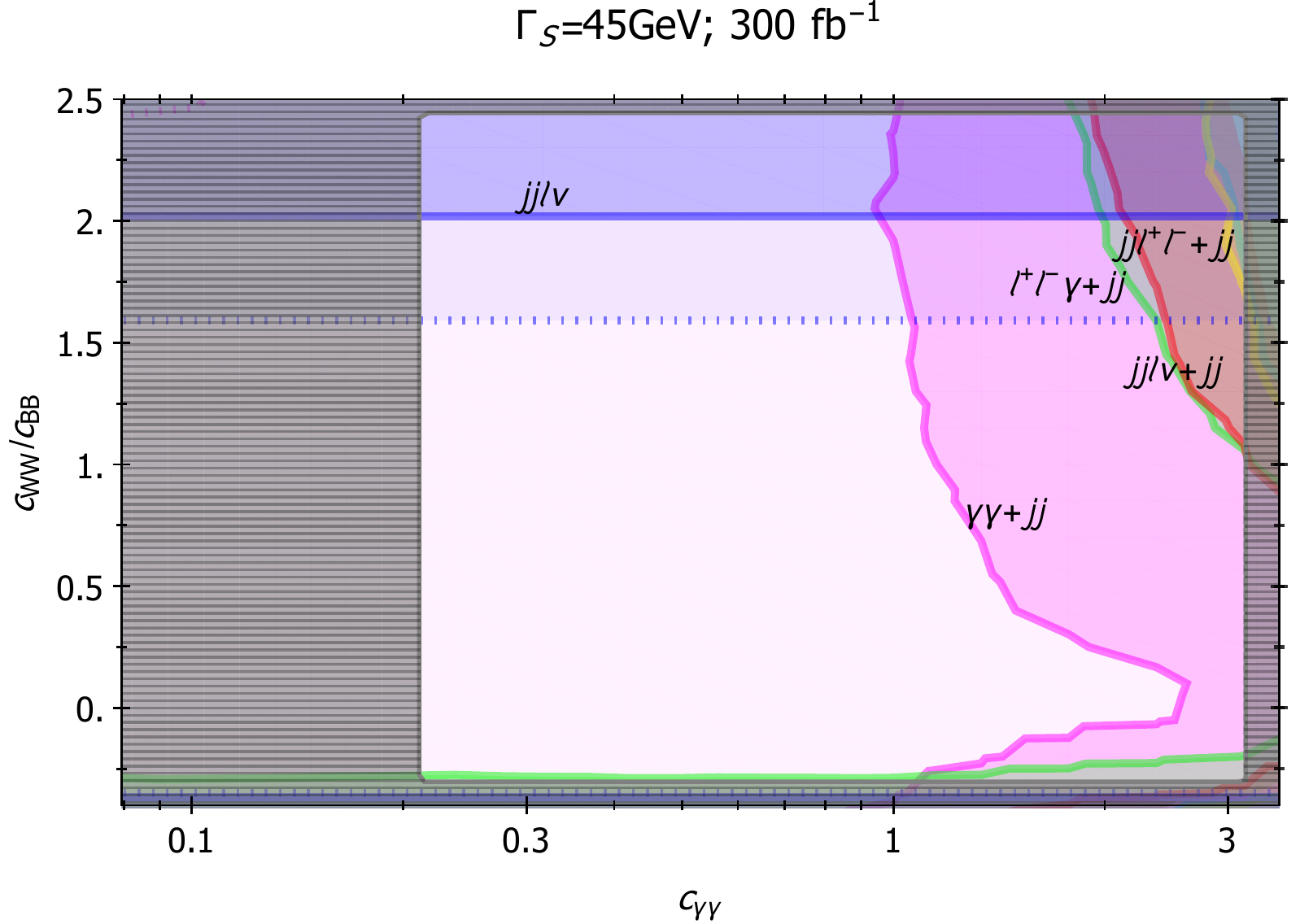}}
 \caption{Parameter space regions for which the CP odd and even hypothesis can be disentangled at the $2\sigma$ level with 300\,fb$^{-1}$. The region where the CP nature can be determined by the different channels is given by the area above the corresponding line. The grey striped regions are excluded (see Fig.~\ref{fig:parameterspace}). In the left panel we assume no extra contributions to $\Gamma_S$, while in the right panel we fix $\Gamma_S = 45$\,GeV. The light area enclosed by the dashed lines stands for the 1.7$\sigma$ region.}\label{fig:final}
\end{figure*}
with $\theta$ the angle between the $p_T$ of the two tagged forward jets. This observable has been previously considered in the literature in the context of Higgs studies (see for example~\cite{Englert:2012xt,Djouadi:2013yb}). As a matter of example, we show the distribution for $\theta^\text{VBF}$ as reconstructed in four-lepton signal events in Fig.~\ref{fig:dvbf}. As in the GF case, the distribution in other channels does not present significant differences. The discrimination power of this angle is apparent from the plot. In order to quantify it for a given number $N_\text{obs}$ of observed events we proceed as in the previous section. Figure~\ref{fig:davbf} shows the distribution followed by $\mathcal{A}^\text{VBF}$ for $N_\text{obs} = 40$ and $S^\text{QCD} = S^\text{EW}$, under the assumption that the background is negligible. In the $2j 2\gamma$ channel, this number of events can be reached with  luminosities of order $\mathcal{L} \sim 60$\,fb$^{-1}$ for $(c_{\gamma\gamma},c_{WW}/c_{BB}) = (2,2)$, while $(c_{\gamma\gamma},c_{WW}/c_{BB}) = (0.1,0.1)$ requiress $\mathcal{L}\sim 200$\,fb$^{-1}$. We plot the one sigma statistical interval as a function of the total number of observed events in Fig.~\ref{fig:resvbf} under the assumption of no background (left panel) and as much background as signal (right panel).
It turns out that less than 40 (60) events are necessary to start disentangling the CP properties of $S$ if there is no background (if there is as much background as signal). Despite this result being apparently much better than the one obtained in GF (see Fig.~\ref{fig:resgf}), in practice VBF is much suppressed (see Eq.~\ref{eq:gfx} and Fig.~\ref{fig:vbfxsecs}) and they are hence complementary.\vspace{-0.2cm}
\section{Results}
\label{sec:results}
%

%
For each point in the parameter space region and for each of the eight event categories $i$ defined for GF and VBF, we compute $\mathcal{A}^\text{GF}$ and $\mathcal{A}^\text{VBF}$ by estimating the number of signal and background events. For a fixed luminosity, the latter can be derived from Tables~\ref{tab:effgf} and~\ref{tab:effvbf}. The number of signal events in each case can be in turn computed as
\begin{equation}
 N_\text{signal} = \sum_i\sigma\times \text{BR}\left(S\rightarrow i\right)\times \epsilon_i,
\end{equation}
where $\epsilon_i$ stands for the corresponding experimental efficiency as provided in Tables~\ref{tab:effgf} and~\ref{tab:effvbf}, too. We assume these efficiencies to be independent of the coefficients of the operators in Eq.~\ref{eq:par}. We have checked that this is the case in almost the whole parameter space, small variations arising only in the VBF $2\gamma\,2j$ channel for $c_{WW}\ll c_{\gamma\gamma}$. At any rate, this region is dominated by $S^\text{QCD}$ and therefore not sensitive to these variations.
Figure~\ref{fig:final} shows the regions where,  with a total luminosity of 300\,fb$^{-1}$,  the CP-odd hypothesis can be excluded at the $2\sigma$ level in favor of the CP-even using the two asymmetries separately (in the left panel no extra sources for $\Gamma_S$ are considered while in the right panel $\Gamma_S = 45$\,GeV instead). These regions are defined by requiring the mean value of $\mathcal{A}$ in the odd case to be separated by at least 2$\sigma$ from the mean value of $\mathcal{A}$ in the even case. For the matter of an example, this separation ($d$) is also shown in  Figs.~\ref{fig:dagf} and ~\ref{fig:davbf}.

The separation between the two hypothesis exceeds 1.5$\sigma$ throughout the parameter space. It is important to mention that small variations on the efficiency and cross section of the $2\gamma\, 2j$ channel can make the corresponding region to look notably different \textit{for a fixed luminosity}. In that respect, an optimisation of the different cuts, as well as other more sophisticated analyses like the matrix-element method used~\cite{Chen:2014pia} in the four-lepton Higgs decay channel, can help covering larger regions of the parameter space. At any rate, even with the basic cuts used in our analysis, luminosities slightly larger than 300\,fb${}^{-1}$ will be sufficient to test  at the 2$\sigma$ confidence level the whole parameter space compatible with 8\,TeV constraints and 13\,TeV data.

The area excluded by searches at 8\,TeV (see Fig.~\ref{fig:parameterspace}) has been superimposed. Note that this area would be  smaller if the required diphoton cross section at 13\,TeV was smaller than the 8\,fb that we are using throughout this letter. With 30\,fb$^{-1}$ only a small portion of the available parameter space can be tested.

It can be shown that GF and VBF channels are complementary, the former being mostly sensitive to the upper region with even small $c_{\gamma\gamma}$. It is also worth emphasizing the role played by semileptonic $W^+ W^-$ decays. This is in contrast with Higgs physics, for which considering these final states is not even possible, inasmuch as the signal peaks in the region populated by the huge $W^\pm +$ jets background. At any rate, the dominant channel is given by $S\rightarrow \gamma\gamma$ in VBF. Indeed, the fact that gluon-initiated processes can also contaminate the EW VBF selection makes this channel sensitive even to regions of small EW couplings, which require large $c_{gg}$.\vspace{-0.2cm}

\section{Conclusions}
\label{sec:conclusions}
The recently observed diphoton excess around $750$\,GeV is triggering a lot of attention. One of the most widely studied explanations relies on a spin-0 real singlet with effective interactions to the SM gauge bosons. In this letter we have thus adopted this setup and discussed the LHC reach for unraveling the CP properties of such a resonance. First, we reviewed the current constraints and commented on the possibility of avoiding flat directions (e.g. resolving the individual couplings to photons and gluons) by considering the associated production with a gauge boson. We have then studied the LHC potential for unraveling the CP nature of such a scalar. Two different asymmetries have been covered in this regard. These are to be constructed out of events produced in gluon and vector-boson fusion respectively. We have shown that events in these categories can be efficiently tagged at the experimental level while keeping backgrounds under control. We have emphasized that as few as $\sim 50$ events are needed to separate the CP even and odd hypotheses. This number of events can be reached in different regions of the parameter space during the next LHC run. In particular, for the full run all the parameter space region is expected to be probed, relying mainly on the VBF $2\gamma\, 2j$ channel.

\section*{Acknowledgements}
We are particularly grateful to Pedro Schwaller for  enlightening discussions at the beginning of this project and for useful comments on the manuscript. We are also thankful to Kai Schmidt-Hoberg, Daniel Stolarski, Alfredo Urbano and Roberto Vega-Morales for valuable feedback. C.G. is supported by the European Commission through the Marie Curie Career Integration Grant 631962 and by the Helmholtz Association. M.R. is supported by la Caixa, Severo Ochoa grant program. M.R. and T.V. are supported by the Spanish Ministry MEC under grants FPA2014-55613-P and FPA2011-25948, by the Generalitat de Catalunya grant 2014-SGR-1450, by the Severo Ochoa excellence program of MINECO (grant SO-2012-0234). 

%

\end{document}